\documentclass[cameraready]{Interspeech}

\title{BiMamba2 Masked Discrete-Unit Prediction for Multilingual Speech Representation for Unsupervised Speech in the Wild Challenge}

\author[affiliation={1}]{Prakriti}{Subedi}
\author[affiliation={1}]{Howard}{Prioleau}
\author[affiliation={1}]{Saurav}{Aryal}

\address{$^1$ Howard University, Washington D.C., USA}

\email{prakriti.subedi@bison.howard.edu, howard.prioleau@bison.howard.edu, saurav.aryal@howard.edu}


\usepackage{booktabs}
\usepackage{siunitx}
\usepackage{amsmath}
\usepackage{url}
\usepackage{graphicx}
\usepackage{tikz}
\usepackage{svg}
\usepackage{float}
\usetikzlibrary{arrows.meta,positioning,calc,fit}

\begin{document}
\maketitle

% ============================================================
% ABSTRACT
% No LaTeX commands. No citations. Max 1000 characters incl. spaces.
% Current count: ~940 characters. Verify before submission.
% ============================================================
\begin{abstract}
We describe our submission to the Unsupervised Speech in the Wild (UPS) Challenge at Interspeech 2026, a bidirectional Mamba-2 (BiMamba2) encoder trained with masked discrete-unit prediction following the HuBERT-style paradigm. The 47.88M-parameter model is trained on 250 hours of speech across 67 languages from the MLCommons Unsupervised People's Speech dataset, with no labeled data. The objective combines masked k-means pseudo-label prediction with language identification supervision and VICReg regularization. On official evaluation, the system achieves an Adjusted Rand Index of 0.735, exceeding four baselines on speaker clustering. Language identification macro-F1 (0.073) and character error rate (0.870) remain below supervised baselines. We analyze a local–official discrepancy in metric scale and checkpoint ranking, highlighting limitations of in-distribution diagnostics for predicting Dynabench probe outcomes.
\end{abstract}

% ============================================================
\section{Introduction}
\label{sec:intro}
% ============================================================

Learning robust speech representations from unlabeled, acoustically diverse data remains a central challenge in the field. While supervised pretraining on large, carefully curated corpora has driven major advances in high-resource settings~\cite{baevski2020wav2vec,hsu2021hubert}, most of the world’s languages lack the labeled data required to benefit from such approaches~\cite{ardila2020commonvoice,conneau2020xlsr}. As a result, existing speech models exhibit stark performance disparities across languages and acoustic conditions, limiting their real-world applicability. This imbalance raises fundamental questions about how to build speech representations that generalize beyond a narrow set of well-resourced scenarios. Similar to text, self-supervised methods that learn directly from raw audio offer a more scalable path toward broad-coverage multilingual representation~\cite{yang2021superb,shi23g_interspeech,baevski2022data2vec}.

The Unsupervised Speech in the Wild (UPS) Challenge~\cite{ups2026} formalizes this setting: models are trained entirely without labels on the MLCommons Unsupervised People's Speech dataset~\cite{mlcommons_ups,galvez2021people} and evaluated on three probe tasks --- language identification (macro-F1), automatic speech recognition (CER), and speaker clustering (ARI) --- applied to held-out audio without task-specific fine-tuning.

Our submission applies masked discrete-unit prediction in the HuBERT style~\cite{hsu2021hubert} to a bidirectional adaptation of the Mamba2 state-space model (BiMamba2)~\cite{dao2024mamba2} for the Open Filtering Sub-Track. We chose Mamba for its linear-time modeling of sequential signals~\cite{gu2023mamba}, and adopted a bidirectional formulation because the challenge evaluates frozen representations where access to both past and future context yields richer embeddings for non-causal probes. BiMamba2 was preferred over bidirectional alternatives such as Hydra~\cite{hwang2024hydra} and Vision Mamba~\cite{zhu2024vision} for its leaner design; eliminating sequence shifting and leveraging the optimized Mamba2 kernel~\cite{dao2024mamba2} for faster throughput. The masked objective predicts discrete cluster targets~\cite{oord2017vqvae} at masked positions following the BERT paradigm~\cite{devlin2019bert} adapted for speech~\cite{hsu2021hubert,chen2022wavlm}, supplemented by VICReg regularization~\cite{bardes2022vicreg} and a language identification loss. The model is trained on a curated $\sim$250-hour, 67-language subset of the MLCommons UPS dataset using a shard-based pipeline with precomputed log-mel features, VAD filtering, and language-aware sampling.

The contributions of this paper are: (i)~a BiMamba2-based masked discrete-unit model for multilingual unsupervised speech representation; (ii)~a shard-based multilingual training pipeline with VAD filtering and language-aware sampling and (iii)~a bidirectional evaluation mismatch analysis with reproducibility implications for future challenge participants.

% ============================================================
\section{Data and Preprocessing}
\label{sec:data}
% ============================================================

\subsection{Dataset}

All training data is drawn from the MLCommons Unsupervised People's Speech dataset~\cite{mlcommons_ups,galvez2021people}, the official challenge source~\cite{ups2026}. No external data is used. After filtering, the corpus comprises approximately 87{,}400 utterance chunks across 438 precomputed shards, covering roughly 250 hours (approximately 100 hours English, 150 hours non-English) in 67 languages (Table~\ref{tab:langs}). VAD timestamps and language labels are consumed directly from MLCommons precomputed metadata (\texttt{vad\_results.jsonl}, \texttt{lang\_id\_results.jsonl}); no local VAD or LID inference is run.

\begin{table}[th]
  \caption{Training corpus languages (67 total, language codes from UPS metadata),
  ordered by utterance count descending.}
  \label{tab:langs}
  \centering\small
  \begin{tabular}{p{7.2cm}}
    \toprule
    en, es, de, pt, ar, fr, it, gl, ne, id, pl, zh, nl, ca, sv,
    am, ru, hu, hi, eu, ur, jw, ms, tr, ko, el, kn, sa, cy, ta,
    sw, uk, bs, ro, si, fa, te, sr, so, hr, ja, nn, la, da, tl,
    hy, vi, sq, pa, sk, ka, yo, be, bn, he, bg, km, br, af, sd,
    lt, sn, cs, mr, ps, th, ml \\
    \bottomrule
  \end{tabular}
\end{table}

\subsection{Filtering and Feature Extraction}

Clips are retained if: (i)~the VAD speech-to-clip ratio is
$\geq\!0.30$; (ii)~the longest VAD segment is $\geq\!10$~s;
(iii)~total extractable speech across 10~s-aligned chunks is
$\geq\!30$~s. Entries flagged \texttt{nospeech}, missing
timestamps, or exceeding 200 clips per tar archive are discarded.
At dataloader time, an additional filter requires speech ratio
$\geq\!0.25$ and at least one VAD segment long enough for a full
10~s chunk at 16~kHz.

Log-mel spectrograms are computed offline into precomputed shards
(up to 200 examples each, stored as \texttt{.pt} files),
decoupling feature extraction from model training. Features use
$n_\text{mels}=80$, FFT size 400, hop length 160 samples
(\SI{10}{\milli\second} frame shift), 16~kHz sample rate; each
chunk is fixed at 10~s (1001 frames).

\subsection{Language-Aware Batch Sampling}

English comprises approximately 40\% of utterances after
filtering. To mitigate this dominance, a hybrid sampling strategy
draws 70\% of each batch from a language-balanced sampler and
30\% from uniform shard sampling, with a hard cap of 10\% English
per batch. Shards and within-shard order are reshuffled each
epoch.

% ============================================================
\section{Model Architecture}
\label{sec:model}
% ============================================================

\subsection{Overview}
The model follows a masked discrete-unit prediction framework inspired by HuBERT~\cite{hsu2021hubert}, chosen for its ability to learn strong speech representations without labeled data by predicting self-discovered discrete targets. A stack of BiMamba2 layers~\cite{dao2024mamba2} serves as the encoder backbone, providing bidirectional context with linear-time complexity. $k$-means cluster pseudo-labels ($k=200$, computed from raw log-mel frames) provide the prediction targets at masked positions~\cite{oord2017vqvae,chen2023beats}, removing the need for an external tokenizer. Two auxiliary objectives supplement the masked prediction loss: VICReg regularization~\cite{bardes2022vicreg} encourages informative, decorrelated embeddings, while a language identification loss injects multilingual structure into the representation space. Both operate on mean-pooled hidden states. No separate decoder is used; a linear head projects encoder outputs directly to pseudo-label logits.

% ============================================================
% FIGURE: BiMamba2 block diagram
% Replace the \fbox placeholder with \includegraphics before
% submission. Suggested content: forward pass showing
%   in_proj → split(z, xBC, dt) → conv1d+SiLU → split(x,B,C)
%   → bimamba_chunk_scan → RMSNorm-gate → out_proj
%   with pre-LN and residual wrapper.
% ============================================================

\subsection{BiMamba2 Encoder Layer}
\begin{figure}[H]
  \centering
  \includegraphics[width=.65\columnwidth]{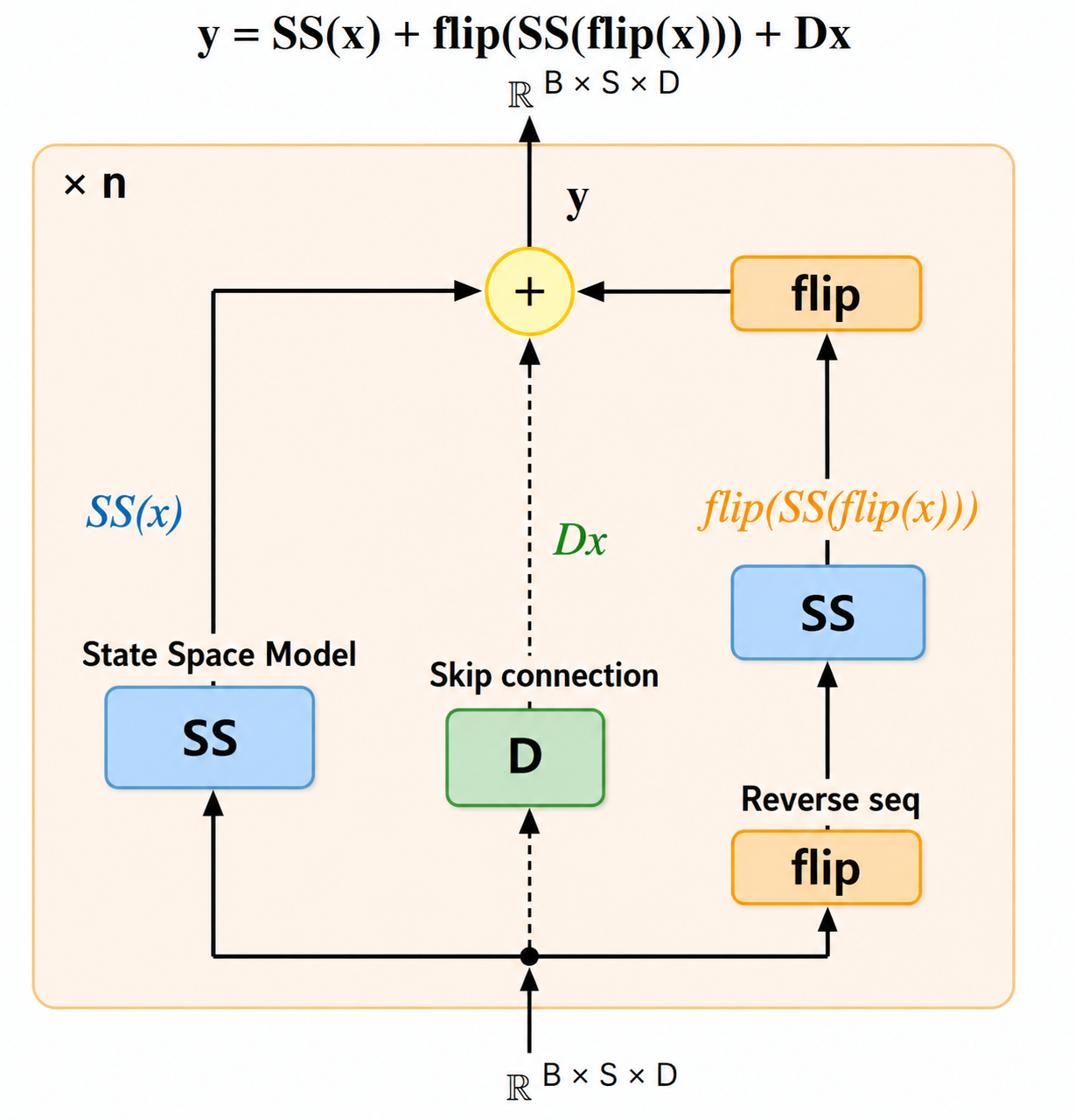}
\caption{\textbf{BiMamba2 encoder layer} (repeated $n$ times). Given
$\mathbf{x}\in\mathbb{R}^{B\times S\times D}$, the layer computes
$\mathbf{y}=\mathrm{SS}(\mathbf{x})+\mathrm{flip}(\mathrm{SS}(\mathrm{flip}(\mathbf{x})))+D\mathbf{x}$,
where $\mathrm{SS}$ is the Mamba2 state-space model and $\mathrm{flip}$ reverses
the sequence. The forward and backward SS paths provide a bidirectional
representation, while $D\mathbf{x}$ acts as a linear skip connection.}
  \label{fig:bimamba2}
\end{figure}

Each BiMamba2 layer processes an input $\mathbf{x}\in\mathbb{R}^{B\times S\times D}$ through three parallel paths whose outputs are summed element-wise. The \emph{forward path} applies the Mamba2 state-space model $\mathrm{SS}$ directly to the input sequence, capturing left-to-right temporal dependencies. The \emph{backward path} first reverses the sequence along the time axis via $\mathrm{flip}$, applies the same $\mathrm{SS}$ operator, and reverses the output back to the original ordering, thereby capturing right-to-left context. The \emph{skip path} applies a learnable diagonal matrix $D$ to the input, providing a direct linear residual that preserves the original signal and stabilizes gradient flow. Together, these three paths yield the layer output:
\begin{equation}
  \mathbf{y} = \mathrm{SS}(\mathbf{x}) + \mathrm{flip}\big(\mathrm{SS}(\mathrm{flip}(\mathbf{x}))\big) + D\mathbf{x}.
  \label{eq:bimamba2}
\end{equation}

The Mamba2 state-space model $\mathrm{SS}$ within each path follows the structured state-space duality (SSD) framework~\cite{dao2024mamba2}. The input is first linearly projected from $d_{\text{model}}$ to an expanded inner dimension $d_{\text{inner}}$, then passed through a short 1-D depthwise convolution of width $d_{\text{conv}}$ to inject local context. A discretized state-space recurrence with state dimension $d_{\text{state}}$ then selectively compresses and propagates information along the sequence. The recurrence output is normalized with RMSNorm~\cite{zhang2019rmsnorm}, modulated by a multiplicative gate $z$, and projected back to $d_{\text{model}}$. Each full block wraps this computation with pre-layer normalization and an additive residual connection around the entire BiMamba2 layer.

The submitted model uses $d_{\text{model}}=768$, 12 layers, $d_{\text{state}}=16$, $d_{\text{conv}}=7$, expand$\,=2$ ($d_{\text{inner}}=1536$): \textbf{47.88M parameters}.

\subsection{Masking and Pseudo-Labels}

Masking is applied only to valid (non-padding) frames. A total
of $\max(1,\lfloor 0.75 \cdot L_\text{valid}\rfloor)$ frames are
masked using 3--5 contiguous block spans; masked frames are
replaced with a learned embedding. Pseudo-labels are obtained
by offline MiniBatchKMeans~\cite{pedregosa2011sklearn} ($k=200$,
default initialization) applied to per-dimension scaled 80-bin
log-mel frame vectors; no pretrained model is used in this step.

\subsection{Training Objectives}

The primary loss is cross-entropy over $k$-means pseudo-labels
at masked positions~\cite{hsu2021hubert,devlin2019bert}:
\begin{equation}
  \mathcal{L}_\text{mask} = -\frac{1}{|\mathcal{M}|}
  \sum_{t \in \mathcal{M}} \log p_\theta(c_t \mid \tilde{x}),
\end{equation}
where $\mathcal{M}$ is the masked frame set, $c_t$ the pseudo-label
at frame $t$, and $\tilde{x}$ the masked input. VICReg variance
($\lambda_\text{var}=5.0$) and covariance ($\lambda_\text{cov}=1.0$)
penalties~\cite{bardes2022vicreg,zbontar2021barlowtwins} on mean-pooled hidden states
prevent representation collapse. An auxiliary LID cross-entropy
loss on mean-pooled states, using precomputed UPS language labels,
contributes a language-discriminative signal with weight
$\lambda_\text{lid}=0.05$. The total loss is:
\begin{equation}
  \mathcal{L} = \mathcal{L}_\text{mask}
              + \mathcal{L}_\text{VICReg}
              + \lambda_\text{lid}\,\mathcal{L}_\text{lid}.
\end{equation}

% ============================================================
\section{Training}
\label{sec:training}
% ============================================================

The model is optimized with AdamW~\cite{loshchilov2019adamw}
(lr$=10^{-4}$, gradient clip max-norm 0.5). The schedule uses
linear warmup over 2{,}000 steps then cosine decay to $0.1
\times \text{base lr}$. Training uses bfloat16 mixed precision,
batch size 128, and seed 42, with no gradient accumulation.
Training resumed from a step-10{,}500 checkpoint of the same
configuration and ran to step 48{,}000, with checkpoints and
holdout validation every 1{,}500 steps. Hardware: single NVIDIA
RTX~A6000 GPU. Input frames are clamped to $[-20, 20]$; hidden
states pass through \texttt{nan\_to\_num}; pooled VICReg
embeddings are clamped to $[-100, 100]$ for numerical stability.

% ============================================================
\section{Challenge Submission}
\label{sec:submission}
% ============================================================

The submission implements the Dynabench \texttt{ModelController} API~\cite{ups2026}.
Frame-level encoder outputs $[T,d_\text{model}]$ from the post-final-LN encoder are
averaged over three non-overlapping temporal segments (start, middle, end) and
L2-normalized for embedding generation. The audio frontend resamples to 16~kHz,
extracts 80-bin log-mel features (FFT 400, hop 160), and pads or crops to a fixed
10~s window. Checkpoint loading uses conservative fallbacks for compatibility
across training and evaluation runtimes. The BiMamba2 backbone requires its CUDA backend for inference.

% ============================================================
\section{Experimental Setup}
\label{sec:setup}
% ============================================================

\subsection{Official evaluation.}
The UPS Challenge evaluates submitted embeddings on three
downstream probe tasks via Dynabench~\cite{ups2026} without
participant access to probe training details or test labels:
language identification (macro-F1, $\uparrow$), speech
recognition (CER, $\downarrow$), and speaker clustering
(ARI, $\uparrow$). All official scores are from the final
patched submissions. Challenge baselines are
Whisper~\cite{radford2022whisper}, HuBERT-large~\cite{hsu2021hubert},
XLSR~\cite{conneau2020xlsr}, and wav2vec~2.0~\cite{baevski2020wav2vec}.

\subsection{Smaller model configuration.}
To examine the effect of scale, we also submitted a smaller
model ($d_\text{model}=512$, 8 layers, batch size 64, step
49,000) using the same training objectives and pipeline as
the primary model.

\subsection{Local evaluation.}
Two local evaluation protocols were used during development, both on a 45-language split drawn from the same UPS source as the training data. \textit{Holdout validation} (6,646 samples) was run at each checkpoint interval. \textit{Linear-probe evaluation} was run post-hoc on precomputed embeddings using a trained probe~\cite{pedregosa2011sklearn}. The local evaluation split was formed by holding out entire tar shards that were excluded from the training tar list. Because recordings are uniquely assigned to a single tar shard in UPS, this shard-level exclusion guarantees zero recording overlap between training and evaluation; therefore, no additional recording-level deduplication was necessary. Both metrics are reported for analysis in Section~\ref{sec:mismatch} and are not valid proxies for official challenge performance.

% ============================================================
\section{Results}
\label{sec:results}
% ============================================================

\subsection{Official Challenge Results}

Table~\ref{tab:official} reports official Dynabench results.
We submitted two checkpoints for the larger model: step 19,500,
submitted during training as an intermediate evaluation, and
step 48,000, submitted after training completion. Official
evaluation showed step 19,500 to be superior on all three
metrics, and it is designated as our primary result.

The step-19,500 checkpoint achieves ARI \textbf{0.735},
macro-F1 0.073, and CER 0.870. ARI exceeds all four baselines,
by 0.105 over wav2vec~2.0 and 0.635 over XLSR. Macro-F1 is
substantially below Whisper (0.950) and HuBERT-large (0.690),
and CER falls between XLSR (0.900) and HuBERT-large (0.570).

The step-48,000 checkpoint yields similar macro-F1 (0.070) and
ARI (0.710) but markedly higher CER (0.998), indicating
late-stage degradation in phonetic generalization
(Section~\ref{sec:decay}). The smaller model (d=512, 8L, step
49,000) performs substantially below the primary model on all
three tasks.

\begin{table}[th]
  \caption{Official UPS Challenge results (Dynabench).
  $\uparrow$: higher is better; $\downarrow$: lower is better.
  Best per metric in \textbf{bold}.
  $\dagger$: $d{=}512$, 8 layers, batch 64, step 49{,}000.}
  \label{tab:official}
  \centering
  \begin{tabular}{lccc}
    \toprule
    \textbf{System}
      & \textbf{Macro-F1}$\uparrow$
      & \textbf{CER}$\downarrow$
      & \textbf{ARI}$\uparrow$ \\
    \midrule
    Whisper~\cite{radford2022whisper}
      & \textbf{0.950} & 0.790          & 0.560 \\
    HuBERT-large~\cite{hsu2021hubert}
      & 0.690          & \textbf{0.570} & 0.600 \\
    XLSR~\cite{conneau2020xlsr}
      & 0.260          & 0.900          & 0.100 \\
    wav2vec~2.0~\cite{baevski2020wav2vec}
      & 0.250          & 0.980          & 0.630 \\
    \midrule
    Ours (d=512, 8L)$^\dagger$
      & 0.052 & 0.996 & 0.291 \\
    Ours (d=768, 12L, step 48k)
      & 0.070 & 0.998 & 0.710 \\
    \textbf{Ours (d=768, 12L, step 19.5k)}
      & 0.073 & 0.870 & \textbf{0.735} \\
    \bottomrule
  \end{tabular}
\end{table}

\subsection{Local Diagnostic Evaluation}
\label{sec:local_eval}

Table~\ref{tab:local} reports holdout validation scores at both
checkpoints. These are \emph{not} official challenge results and
are reported here to contextualize the local-vs-official mismatch
analyzed in Section~\ref{sec:mismatch}.

\begin{table}[th]
  \caption{Local holdout validation (45 languages, 6{,}646
  samples). \emph{Not} official challenge scores. Intra/inter-class
  similarity computed as mean cosine similarity over mean-pooled
  embeddings per language. Separability ratio $=$ intra-class
  sim.\ $\div$ inter-class sim.; negative value at step 48k
  arises from negative mean inter-class cosine similarity.}
  \label{tab:local}
  \centering
  \begin{tabular}{lcc}
    \toprule
    \textbf{Metric}
      & \textbf{Step 19.5k}
      & \textbf{Step 48k} \\
    \midrule
    Macro-F1          & 0.342  & 0.395 \\
    ARI               & 0.225  & 0.232 \\
    Intra-class sim.  & 0.786  & 0.747 \\
    Inter-class sim.  & 0.004  & $-$0.002 \\
    Separability ratio & 208.98 & $-$366.84 \\
    \bottomrule
  \end{tabular}
\end{table}

Locally, step 48,000 appears marginally stronger than step 19,500
on macro-F1 (0.395 vs.\ 0.342) and ARI (0.232 vs.\ 0.225), yet
official evaluation shows the reverse: step 19,500 substantially
outperforms step 48,000, particularly on CER (0.870 vs.\ 0.998).
This illustrates that local holdout validation did not reliably
predict which checkpoint would generalize better to the official
evaluation distribution.

At step 48,000, mean inter-class cosine similarity is slightly
negative ($-$0.002), making the separability ratio negative and
large in magnitude ($-$366.84). This indicates that at late-stage
training, embeddings from different language groups are on average
weakly anti-correlated in the learned space. In contrast, at step
19,500, inter-class similarity is small but positive (0.004),
producing a well-defined positive separability ratio (208.98). We
discuss this geometric shift in Section~\ref{sec:decay}.

A separate post-hoc linear-probe evaluation on the same
45-language split produced macro-F1 of 0.915, further inflated
by the expressive capacity of a trained probe on in-distribution
data.

% ============================================================
\section{Discussion}
\label{sec:discussion}
% ============================================================

\subsection{Speaker Clustering vs.\ LID and ASR}

The primary model exceeds all ARI baselines while macro-F1 is
0.073 --- a pattern that holds across both model sizes. One
plausible account is that masked discrete-unit prediction
encourages speaker-discriminative structure through within-utterance
frame recovery: speaker prosody, vocal tract characteristics, and
speaking style may be partially recoverable from local acoustic context.
Reliable LID, by contrast, requires consistent coverage of each
language's phonological inventory in the training distribution.
With approximately 250 hours across 67 languages, generalization
to the challenge test set is unlikely for languages with limited
or absent training representation. The auxiliary LID loss (weight
0.05) provides a weak supervision signal but cannot compensate
for this coverage gap. Similar coverage-performance relationships
have been observed in large-scale SSL models trained on
more balanced multilingual
data~\cite{chen2022wavlm,conneau2022fleurs}. This account is a
post-hoc interpretation; no ablation was conducted to confirm it.

\subsection{Late-Stage Embedding Geometry and CER Degradation}
\label{sec:decay}

Official ARI is relatively stable across checkpoints (0.735
vs.\ 0.710), but CER degrades sharply from 0.870 to 0.998 between
step 19,500 and step 48,000. The local holdout data adds a
geometric perspective: at step 48,000, mean inter-class cosine
similarity shifts from slightly positive (0.004) to slightly
negative ($-$0.002), with intra-class similarity declining
modestly (0.786 $\to$ 0.747). This suggests the embedding space
undergoes a structural reorganization in late training: within-class
embeddings remain coherent but between-class embeddings become
weakly anti-correlated, producing a geometry that is well-separated
locally but may not preserve the phonetic similarity structure
required for downstream CER evaluation.

A principled account is that late-stage training under cosine
decay with fixed $k$-means targets progressively specializes the
encoder toward pseudo-label cluster boundaries at the cost of
broader phonetic generalization~\cite{hsu2021hubert,chen2023beats}.
As the learning rate approaches its minimum ($10^{-5}$), gradient
updates increasingly refine fine-grained cluster distinctions
derived from raw log-mel frames without phonetic annotation, which
may not align with the phonetic similarities probed by CER. Iterative
pseudo-label refinement~\cite{hsu2021hubert,chen2023beats} could
mitigate this effect in future work. We acknowledge this
interpretation is a hypothesis; controlled experiments are needed
to confirm it.

\subsection{Bidirectional Local-vs-Official Mismatch}
\label{sec:mismatch}

Local evaluation diverges from official scoring in both directions.
At step 19,500, local macro-F1 (0.342) and post-hoc probe macro-F1
(0.915) both substantially overestimate official macro-F1 (0.073),
while local ARI (0.225) substantially underestimates official ARI
(0.735). Moreover, the locally-better checkpoint (step 48,000,
local macro-F1: 0.395) performs worse than step 19,500 on all
three official metrics, indicating that local holdout validation
is an unreliable predictor of official ranking. Three factors
contribute.

\textbf{Language overlap.} All 45 local eval languages appear in
the training corpus, including the most frequent training languages
(es, de, pt, fr, ar). LID evaluation on in-distribution languages
produces optimistic scores that do not reflect generalization.

\textbf{Train--eval separation by construction.} Local eval shards were constructed
from tar archives excluded from the training tar list (tracked and retained),
so recording overlap with training shards is not expected.

\textbf{Probe methodology mismatch.} The local ARI is computed
using cosine-similarity-based clustering on mean-pooled embeddings,
while the Dynabench speaker clustering probe uses a different
methodology and evaluation set. The high local separability ratio
at step 19,500 (208.98) indicates well-separated clusters in the
local embedding space, but these clusters correspond to language
groups rather than speakers, which explains the low local ARI and
high local macro-F1 simultaneously.

The practical lesson is that valid local proxies for multilingual
challenge evaluation require: (i)~a held-out language set disjoint
from training; (ii)~an evaluation set disjoint from training by construction
(e.g., tar-level exclusion); and (iii)~probe methodology matched to the
official evaluation.

\subsection{Effect of Model Scale}
Scaling from d=512/8L to d=768/12L shows consistent official
gains: most pronounced in ARI (0.291$\to$0.735) and CER
(0.996$\to$0.870), with marginal improvement in macro-F1
(0.052$\to$0.073), consistent with LID being primarily limited
by data coverage. We note that the two configurations differ in
batch size (64 vs.\ 128) and total training steps, so gains
cannot be attributed to model scale alone; controlled ablations
were not conducted.

\section{Conclusion}
\label{sec:conclusion}
We presented a BiMamba2-based masked discrete-unit prediction model for the UPS Challenge, trained on approximately 250 hours of multilingual speech across 67 languages from the MLCommons UPS dataset. The system combines $k$-means pseudo-label prediction at masked frames with auxiliary LID supervision and VICReg regularization, trained via a shard-based precomputed pipeline with language-aware sampling. Our 47.88M-parameter primary checkpoint achieves an official ARI of 0.735, exceeding all four published baselines for speaker clustering, while macro-F1 (0.073) and CER (0.870) remain below supervised baselines due to training data scale and language coverage limitations. Comparing two submitted checkpoints reveals late-stage CER degradation and an embedding geometry shift consistent with over-specialization toward fixed pseudo-label boundaries under a decayed learning rate, motivating iterative target refinement in future work. We document a bidirectional mismatch between local diagnostic and official evaluation scores --- local evaluation overestimates LID and underestimates speaker clustering quality, and locally superior checkpoints do not reliably correspond to officially superior ones --- offering this as a concrete evaluation protocol lesson for multilingual challenge settings. The study is limited by the absence of component ablations, single-run results without variance estimates, a training corpus substantially smaller than those of the baselines, and pseudo-labels computed once without iterative refinement; addressing these gaps is a priority for future work.

% ============================================================
% ACKNOWLEDGMENTS (camera-ready only; leave blank for review)
% ============================================================
\section{Acknowledgments}
This research was supported by the Office of Naval Research, Department of the Navy (FAIN N00014-22-1-2714), the NIH Common Fund under award 1OT2OD032581-01, and the Amazon Research Award. The views expressed are those of the authors and do not necessarily reflect those of the funders or affiliated organizations.

% ============================================================

% ============================================================

% ============================================================
\bibliographystyle{IEEEtran}
\bibliography{mybib}

\section{Generative AI Use Disclosure}
We used generative AI tools to assist with outlining and improving readability of parts of the manuscript. The authors are fully responsible for the content; all claims, numbers, and citations were checked by the authors. Generative AI was not used to produce the experimental results.

\end{document}